\documentclass[3p,times,twocolumn]{elsarticle}
 \biboptions{comma,sort&compress}
 
\usepackage{graphicx}
\usepackage{amsmath}
\usepackage{here}
\usepackage{ecrc}

\usepackage{stackrel}

\volume{00}

\firstpage{1}

\journalname{Nuclear and Particle Physics Proceedings}

\runauth{}

\jid{nppp}

\jnltitlelogo{Nuclear and Particle Physics Proceedings}

\usepackage{amssymb}

\usepackage[figuresright]{rotating}

\newcommand{\Tr}{{\textrm{Tr}}}

\newcommand{\MeV}{\,{\rm MeV}}
\newcommand{\GeV}{\,{\rm GeV}}
\newcommand{\fm}{\,{\rm fm}}
\newcommand{\QCD}{{\textrm{\scriptsize QCD}}}
\newcommand{\HRG}{{\textrm{\scriptsize HRG}}}
\newcommand{\mmin}{{\textrm{\scriptsize min}}}

\begin{document}

\begin{frontmatter}

\title{ Fluctuations and correlations in thermal QCD
 $^*$}
 \cortext[cor0]{Talk given at 21th International Conference in Quantum Chromodynamics (QCD 18),  2 - 6 july 2018, Montpellier - FR}
 \author[label1,label2]{E. Meg\'{\i}as\fnref{fn1}}
   \fntext[fn1]{Speaker, Corresponding author.}
\ead{emegias@ugr.es}
\address[label1]{Departamento de F\'{\i}sica At\'omica, Molecular y Nuclear and Instituto Carlos I de F\'{\i}sica Te\'orica y Computacional, Universidad de Granada, Avenida de Fuente Nueva s/n, 18071 Granada, Spain}
\address[label2]{Departamento de F\'{\i}sica Te\'orica, Universidad del Pa\'{\i}s Vasco UPV/EHU, Apartado 644, 48080 Bilbao, Spain}

 \author[label1]{E. Ruiz Arriola}
\ead{earriola@ugr.es}

 \author[label1]{L.~L. Salcedo}
    \ead{salcedo@ugr.es}

\pagestyle{myheadings}
\markright{ }
\begin{abstract}
We study the equation of state, fluctuations and static correlators of electric charge, baryon number and strangeness, by considering a realization of the Hadron Resonance Gas model in the light flavor sector of QCD. We emphasize the importance of these observables to study, within this approach, the possible existence of exotic and missing states in the hadron spectrum. Some preliminary results for the baryon spectrum have been obtained within a relativistic quark-diquark model, leading to an overall good agreement with the spectrum obtained with other quark models. Finally, it is conjectured, within the Hadron Resonance Gas approach, the existence of a singularity in the correlators at zero temperature, which turns out to be analogous to the divergence of the partition function at the Hagedorn temperature. 
\end{abstract}

\end{frontmatter}
\section{Introduction}
\label{sec:introduction}

Before the advent of QCD, a major inspiring breakthrough came about when Hagedorn found that purely hadronic matter forming a hadron resonance gas (HRG) has an exponentially growing level density implying a limiting temperature of hadronic matter of about $T_H \approx 150\, \MeV$~\cite{Hagedorn:1965st}. Within a quantum virial expansion this corresponds to a weakly interacting gas of resonances~\cite{Dashen:1969ep}, and the commonly accepted reference for hadronic states is the Particle Data Group (PDG)~\cite{Tanabashi:2018oca}. We thus expect the PDG hadronic states to have a one-to-one correspondence with colour neutral eigenstates of the QCD Hamiltonian, however so far these states are commonly interpreted as mesons and baryons. On this respect, we may conclude that evidence of new colour singlet states in the spectrum, if they exist, should be deduced from a detailed but tricky comparison of the HRG model with results from lattice QCD.

Apart from the Equation of State (EoS) of QCD, there are a number of thermal observables that can be used to study the spectrum of QCD by distinguishing between different flavor sectors; these are the fluctuations~\cite{Bazavov:2012jq} and correlations of conserved charges. Recent lattice studies of fluctuations suggest that they are sensitive probes of deconfinement (see e.g. Ref.~\cite{Ding:2015ona} for a recent review). In this work we study these quantities within the HRG approach, and perform a comparison with recent lattice simulations when available.

\section{Thermodynamics of QCD and hadron spectrum}

In this section we will summarize the main properties of the spectrum of QCD and its relation to the thermodynamics of QCD within the HRG approach.

\subsection{Thermodynamics of QCD}

The partition function of QCD
\begin{equation}
Z_{\QCD} = \Tr \, e^{-H_{\QCD}/T}= \sum_n e^{-E_n/T} \,, \label{eq:Zqcd}
\end{equation}
is the fundamental quantity to study the thermodynamic properties of this theory. Written in terms of the eigenvalues of the QCD Hamiltonian, i.e.~$H_{\QCD} \psi_n = E_n \psi_n$, Eq.~(\ref{eq:Zqcd}) illustrates the relation between the thermodynamics of the confined phase and the spectrum of QCD.
\begin{figure*}[htp]
\centering
 \begin{tabular}{c@{\hspace{2.5em}}c}
 \includegraphics[width=0.43\textwidth]{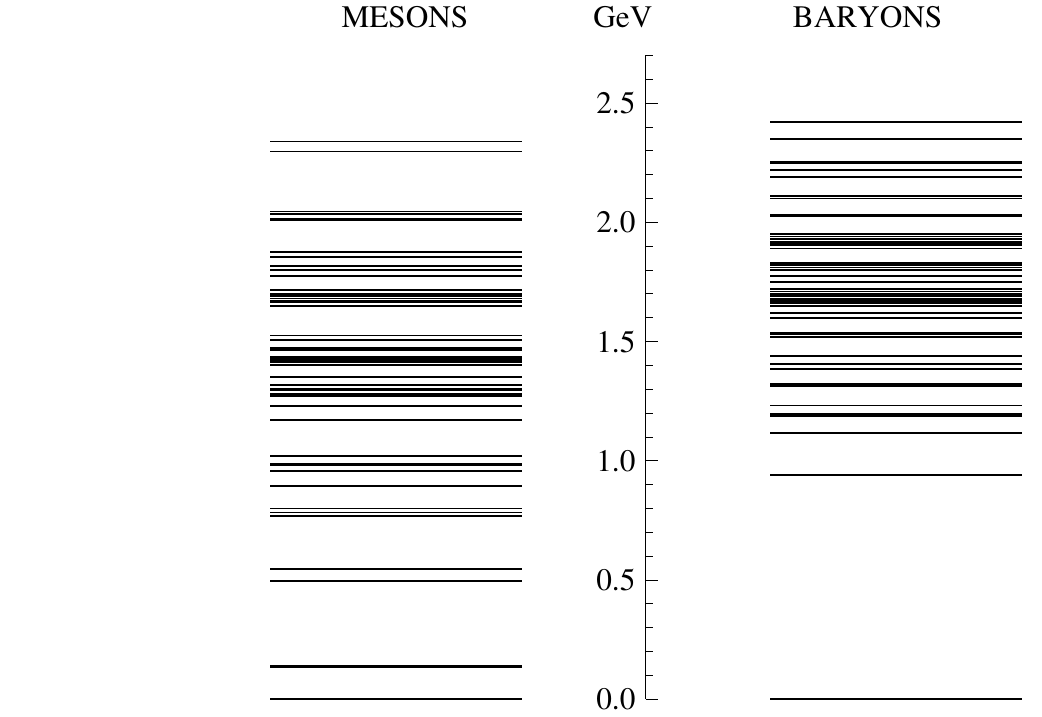} &
 \includegraphics[width=0.43\textwidth]{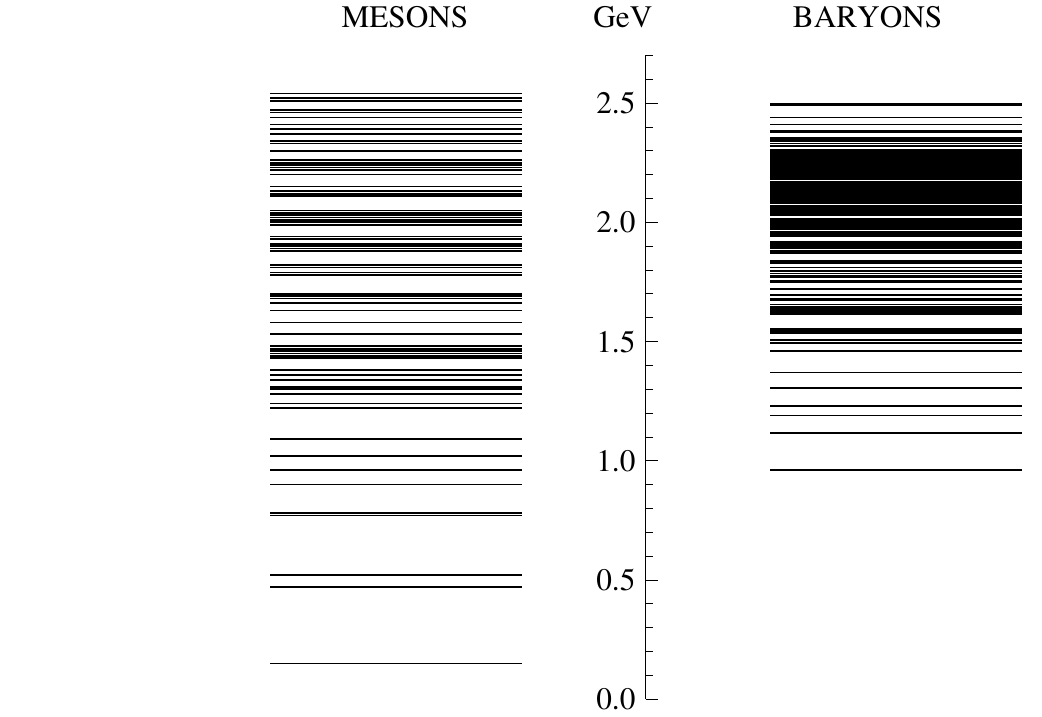} 
\end{tabular}
 \caption{\it Left panel: Mesons and baryons spectrum made of $u$, $d$ and $s$ quarks from the PDG~\cite{Tanabashi:2018oca} (left panel), and from the Relativized Quark Model~\cite{Godfrey:1985xj,Capstick:1986bm} (right panel).}
\label{fig:spectrum}
\end{figure*}

In QCD, the quantized energy levels are the masses of colour singlet states, which are commonly identified with mesons $[q\bar{q}]$ and baryons $[qqq]$. Moreover, it has been conjectured the existence of exotic states, including tetraquarks~$[q\bar{q}q\bar{q}]$, pentaquarks~$[qqqq\bar{q}]$, or hybrid states~($[q\bar{q}g]$ and $[qqqg]$). If these states exist, they would contribute to the partition function. Some microscopic evidence for this comes from recent studies on Polyakov - constituent quarks models, in which it is obtained a low temperature partonic expansion around the vacuum of the form~\cite{Megias:2013xaa,Arriola:2014bfa,RuizArriola:2012wd}:
\begin{equation}
Z_{\QCD} = Z_0 \cdot Z_{[q\bar{q}]} \cdot Z_{[qqq]} \cdot  Z_{[\bar{q}\bar{q}\bar{q}]} \cdot Z_{[q\bar{q}g]} \cdot Z_{[q\bar{q}q\bar{q}]} \cdot  \dots  \,. \label{eq:Z_partonic}
\end{equation}
Subsequent hadronization of these states by using the group properties of the Haar measure and other cluster properties of the Polyakov, leads to a {\it microscopic} derivation of the HRG model. This model was originally proposed by Hagedorn~\cite{Hagedorn:1984hz} under the assumption that physical quantities in the confined phase of QCD admit a representation in terms of hadronic states, which are considered as stable, non-interacting and point-like particles. Within this approach the EoS of QCD is described in terms of a free gas of hadrons~\cite{Hagedorn:1984hz,Tawfik:2004sw}, and the grand-canonical partition function turns out to be
\begin{eqnarray}
\hspace{-1.4cm} &&\frac{1}{V} \log Z_\HRG =  \label{eq:Zhrgm} \\
\hspace{-1.4cm} &&\hspace{0.1cm} = - \int \frac{d^3p}{(2\pi)^3} \hspace{-0.2cm} \sum_{i \in {\rm Hadrons}} \hspace{-0.3cm} \zeta_i g_i \log\left( 1 - \zeta_i e^{-(\sqrt{p^2 + M_i^2} - \sum_a  \mu_a q_a^i )/T} \right) \,, \nonumber
\end{eqnarray}
with~$g_i$ the degeneracy factor, $\zeta_i = \pm 1$ for bosons and fermions respectively, and $M_i$ the mass of the $i$-th hadron. We consider several conserved charges labeled by the index~$a$, with $q_a^i$ the charge of the $i$-th hadron for symmetry $a$, and $\mu_a$ the chemical potential associated to this symmetry. The obvious consequence is that a good understanding of the spectrum of QCD turns out to be crucial for a precise determination of the thermodynamic properties of this theory.

\subsection{Hadron Spectrum}

So far, the states listed by PDG echo the standard quark model classification for mesons~$[q{\bar q}]$ and baryons~$[qqq]$. Then, it would be pertinent to consider also the Relativized Quark Model (RQM) for mesons~\cite{Godfrey:1985xj} and baryons~\cite{Capstick:1986bm}. We show in Fig.~\ref{fig:spectrum} the hadron spectrum with the PDG compilation (left) and the RQM spectrum (right). The comparison clearly shows that there are {\it further states} in the RQM spectrum above some scale $M > M_{\mmin}$ that may or may not be confirmed in the future as mesons or hadrons, although they could also be exotic, glueballs or hybrids.

\begin{figure*}[htb]
\centering
 \begin{tabular}{c@{\hspace{4.5em}}c}
 \includegraphics[width=0.43\textwidth]{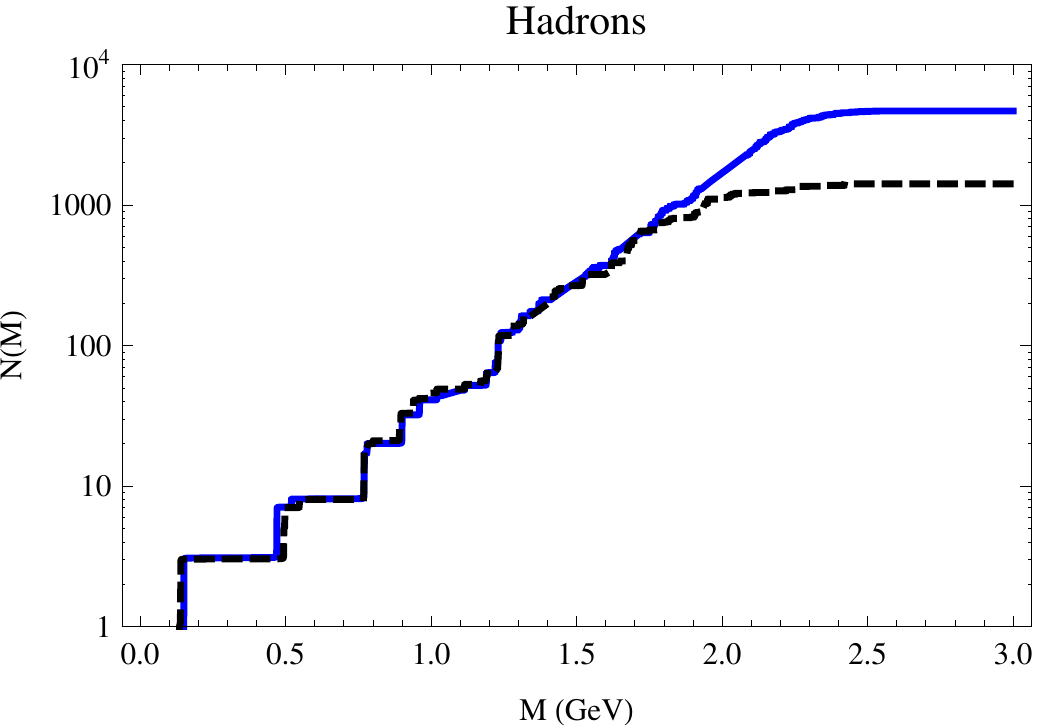} &
\includegraphics[width=0.43\textwidth]{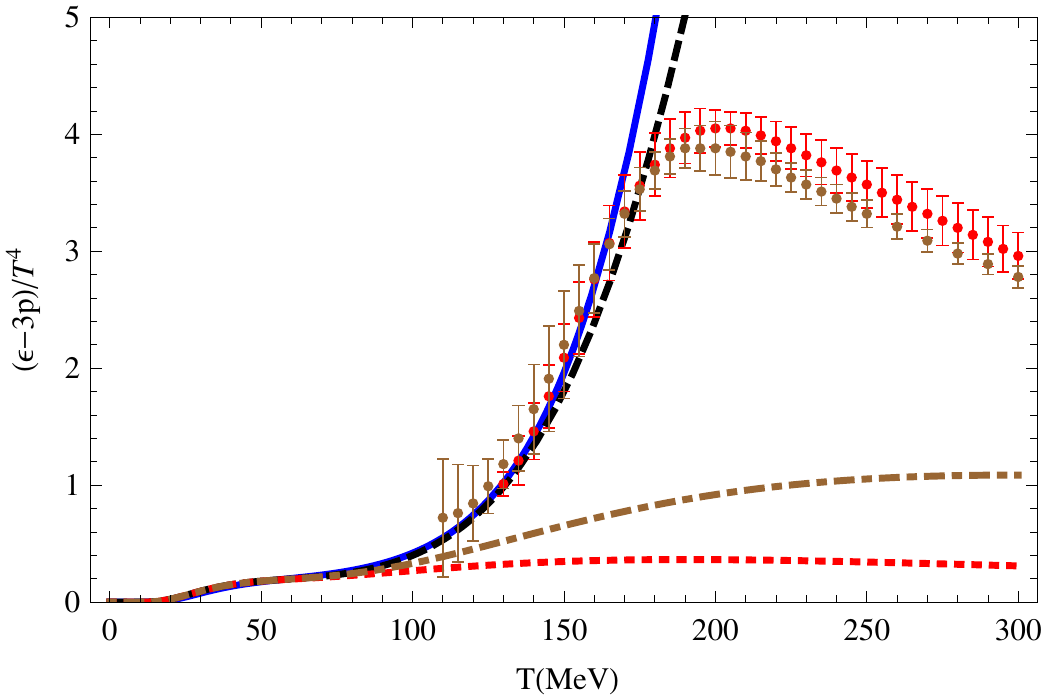} 
\end{tabular}
\vspace{-0.4cm}
 \caption{\it Left panel: Cumulative number of hadrons for the PDG~\cite{Tanabashi:2018oca} (dashed black) and the RQM (solid blue)~\cite{Godfrey:1985xj,Capstick:1986bm}. Right panel: Trace anomaly as a function of temperature in lattice QCD~\cite{Borsanyi:2013bia,Bazavov:2014pvz} vs HRG using PDG (dashed black) and RQM (solid blue) spectra. We also plot just the contribution of states with $M < 0.6 \, \GeV$ (dotted) and $M < 0.8 \, \GeV$ (dotted-dashed).
  }
\label{fig:trace_anomaly}
\end{figure*}

The cumulative number of states is a very useful function for the characterization of the spectrum. It is defined as the number of bound states below some mass~$M$, i.e.
\begin{equation}
N(M) = \sum_i g_i \, \Theta(M-M_i) \,,
\end{equation}
where $\Theta(x)$ is the step function, so that the density of states writes~$\rho(M) = dN(M)/dM$. As it is obvious from the scheme of Eq.~(\ref{eq:Z_partonic}), this function will have contributions from any kind of states, i.e.
\begin{equation}
N(M) = N_{[q{\bar q]}}(M) + N_{[qqq]}(M) + N_{[q{\bar q}q{\bar q}]}(M) +  \cdots \,.
\end{equation}
A derivative expansion~\cite{Caro:1994ht} can be used to evaluate the cumulative number of a Hamiltonian, and this is closely related to a semiclassical expansion. Using these techniques, one can predict that the large mass expansion of these contributions is~$N_{[q{\bar q}]} \sim  M^6$, $N_{[qqq]} \sim M^{12}$, $N_{[q{\bar q}q{\bar q}]} \sim M^{18} \,,$ etc~\cite{Arriola:2014bfa}. This means that each kind of hadron dominates the function $N(M)$ at a different scale. After adding all these contributions, one obtains the conjectured behavior of the Hagedorn spectrum~\cite{Arriola:2013jxa}
\begin{equation}
N_{\HRG}(M) \sim e^{M/T_H} \,.
\end{equation}
This exponential behavior leads to a partition function that becomes divergent at some finite value of the temperature, i.e.
\begin{equation}
Z_\HRG  = \Tr \, e^{-H_{\rm HRG}/T}  \stackrel[T \to T_H^-]{\longrightarrow}{} \frac{A}{T_H - T} \,,  \label{eq:ZHRG}
\end{equation}
where $T_H \approx 150 \,\textrm{MeV}$ is the so-called Hagedorn temperature. We show in Fig.~\ref{fig:trace_anomaly} the trace anomaly computed with Eq.~(\ref{eq:Zhrgm}) by using the PDG and RQM spectrum. One can see that both spectra lead to a good description of lattice data for~$T \lesssim 0.8 T_c$ so that, at least for the EoS, the non-interacting HRG works well in this regime of temperatures.

\subsection{Baryon spectrum from a quark-diquark model}

\begin{figure}[htb]
 \includegraphics[width=0.43\textwidth]{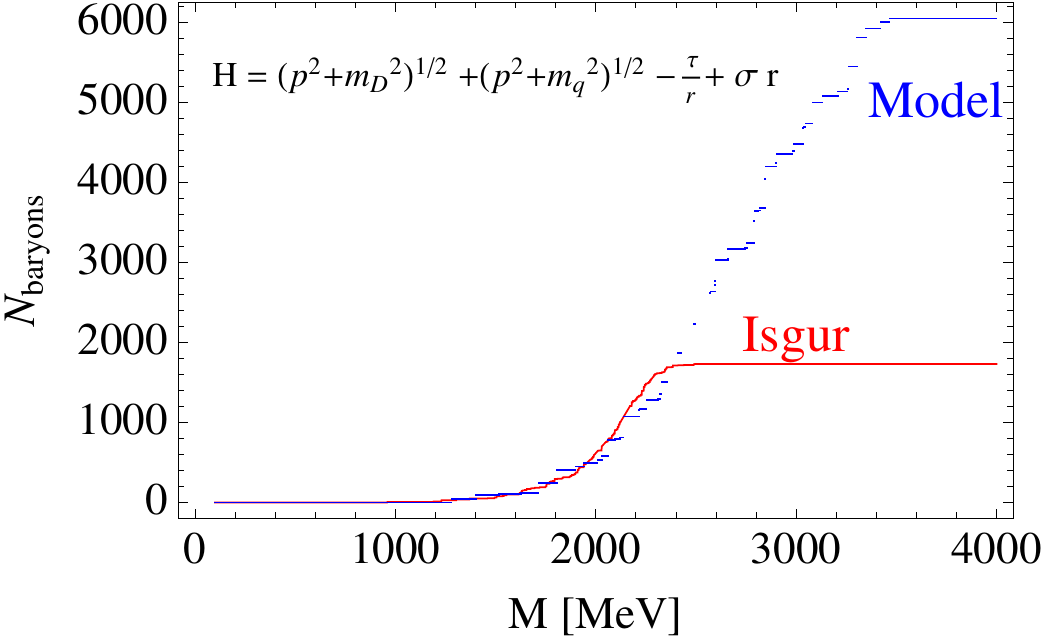}
\vspace{-0.4cm}
 \caption{\it Cumulative number of baryons from the quark-diquark model of Eq.~(\ref{eq:H_qD}) (blue). For comparison, we display as well the cumulative number of baryons from the RQM~\cite{Capstick:1986bm} (red).}
\label{fig:spectrumQD}
\end{figure}

There is nowadays some discussion about the most probable spatial configuration of quarks inside baryons. One interesting possibility would be that they are distributed according to an isosceles triangle. This is the idea behind an easily treatable class of models, the so-called relativistic quark-diquark models, where the baryons are assumed to be composed of a constituent quark, $q$, and a constituent diquark, $D \equiv (qq)$~\cite{Santopinto:2014opa}, and the Hamiltonian writes
\begin{equation}
H_{qD} =  \sqrt{\vec{p}^2 + m_q^2 } + \sqrt{\vec{p}^2 + m_D^2}  + V_{qD}(r) \,.  \label{eq:H_qD}
\end{equation}
We take the quark-diquark potential to be the same as the quark-antiquark potential, i.e.
\begin{equation}
V_{qD}(r) = V_{q{\bar q}}(r) = - \frac{\tau}{r} + \sigma r \,,
\end{equation}
with $\sigma = (0.42\, \GeV)^2$.~\footnote{A theoretical justification of this will be presented in Ref.~\cite{Megias:2018inprogress}.} After considering a convenient choice of the parameters of the model, in particular 
\begin{equation}
\begin{aligned}
&&\hspace{-0.3cm} \tau = \pi/12 \,, \hspace{0.4cm} \qquad m_D = 0.62 \,\GeV \,, \\
&&\hspace{-0.7cm} m_{u,d} = 0.2 \, \GeV \,, \qquad m_s = 0.37 \,\GeV \,,
\end{aligned} \label{eq:param}
\end{equation}
one can diagonalize the Hamiltonian and obtain the spectrum of baryons. We display in Fig.~\ref{fig:spectrumQD} the spectrum of $H_{qD}$, and compare it to the RQM spectrum~\cite{Capstick:1986bm}. It is quite remarkable that below~$M < 2400\, \MeV$  the quark-diquark spectrum turns out to be in good agreement with the RQM spectrum. While the authors of Ref.~\cite{Capstick:1986bm} do not compute baryon masses heavier than this, we have obtained within the present quark-diquark model further states up to~$M \approx 3400 \, \MeV$. These states will contribute to the EoS of QCD as well as other thermal observables. In fact, the choice of parameters of Eq.~(\ref{eq:param}) are motivated by thermal observables. A detailed study of this model and the consequences for the QCD thermodynamics is currently in progress~\cite{Megias:2018inprogress}.

\section{Fluctuations of conserved charges in a thermal medium}

Conserved charges $[Q_a,H]=0$ play a fundamental role in the thermodynamics of QCD. In the ({\it uds}) flavor sector of QCD the conserved charges are the electric charge~$Q$, the baryon number~$B$, and the strangeness~$S$. While their thermal expectation values in the hot vacuum are vanishing, i.e.~$\langle Q_a\rangle_T = 0$ where $Q_a \in \{ Q, B, S \}$, they present statistical fluctuations, characterized by susceptibilities~\cite{Bazavov:2012jq,Bellwied:2015lba,Asakawa:2015ybt}~\footnote{One can also work in the quark-flavor basis, $Q_a \in \{ u, d, s\}$, where $u$, $d$ and $s$ is the number of up, down and strange quarks. In this basis~$B = \frac{1}{3}(u+d+s)$, $Q = \frac{1}{3}(2u-d-s)$ and $S= -s$.}
\begin{equation}
\hspace{-0.5cm} \chi_{ab}(T) \equiv \frac{1}{V T^3}  \langle \Delta Q_a \Delta Q_b \rangle_T \,, \quad \Delta Q_a = Q_a - \langle Q_a \rangle_T \,.
\end{equation}
QCD at high temperature behaves as an ideal gas of quarks and gluons, and the susceptibilities approach in this limit to
\begin{equation}
\left\{ 
\begin{array}{ccc}
\chi_{BB}(T) &\hspace{-0.5cm} \to\hspace{0.1cm} 1/N_c  \nonumber \\
\chi_{QQ}(T) &\hspace{-0.2cm} \to\hspace{0.1cm} \sum_{i=1}^{N_f} q_i^2  \nonumber \\
\chi_{SS}(T) &\hspace{-1.0cm} \to\hspace{0.1cm} 1  \nonumber 
\end{array}\right. \,.
\end{equation}
The susceptibilities can be computed from the grand-canonical partition function by differentiation with respect to the chemical potentials, i.e.
\begin{equation}
\hspace{-0.5cm} -\frac{\partial \Omega}{\partial\mu_a} \Bigg|_{\mu_a = 0} \hspace{-0.3cm} = \langle Q_a \rangle_T \,, \quad  -T \frac{\partial^2 \Omega}{\partial\mu_a \partial\mu_b}  \Bigg|_{\mu_a = 0 = \mu_b} \hspace{-0.8cm}= \langle \Delta Q_a \Delta Q_b \rangle_T \,,  \label{eq:susc}
\end{equation}
where~$\Omega = -T \log Z$ is the thermodynamical potential.

\begin{figure*}[tbh]
\centering
 \begin{tabular}{c@{\hspace{4.5em}}c@{\hspace{4.5em}}c}
 \includegraphics[width=0.26\textwidth]{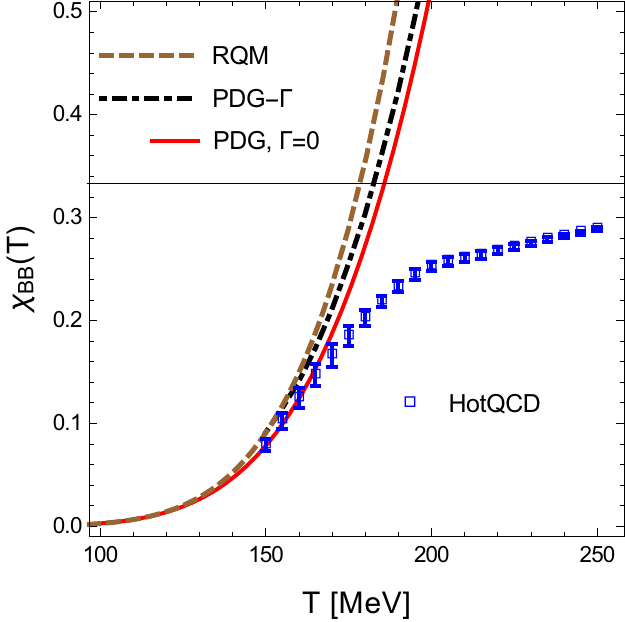} &
 \includegraphics[width=0.26\textwidth]{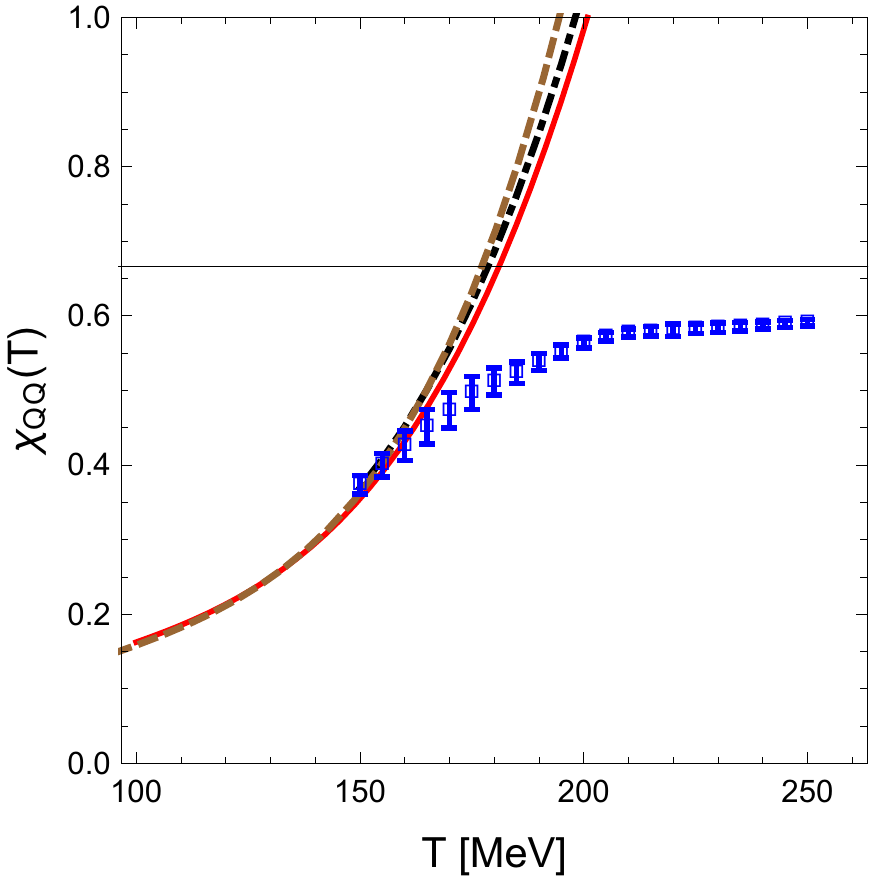} &
 \includegraphics[width=0.26\textwidth]{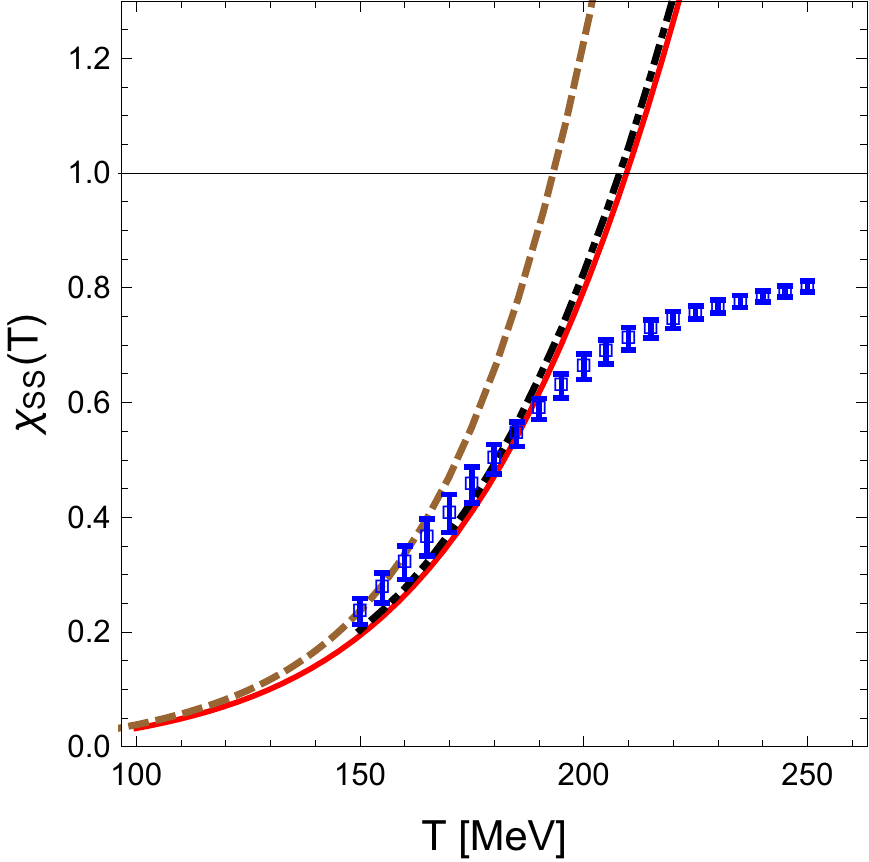} \\
 \includegraphics[width=0.26\textwidth]{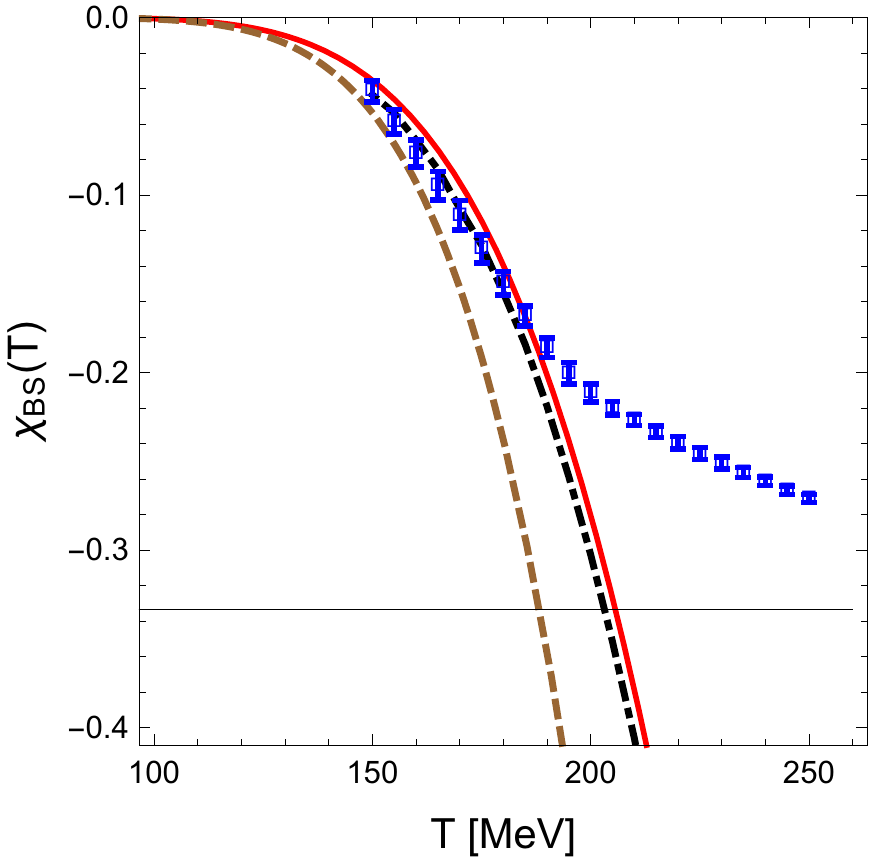} &
 \includegraphics[width=0.26\textwidth]{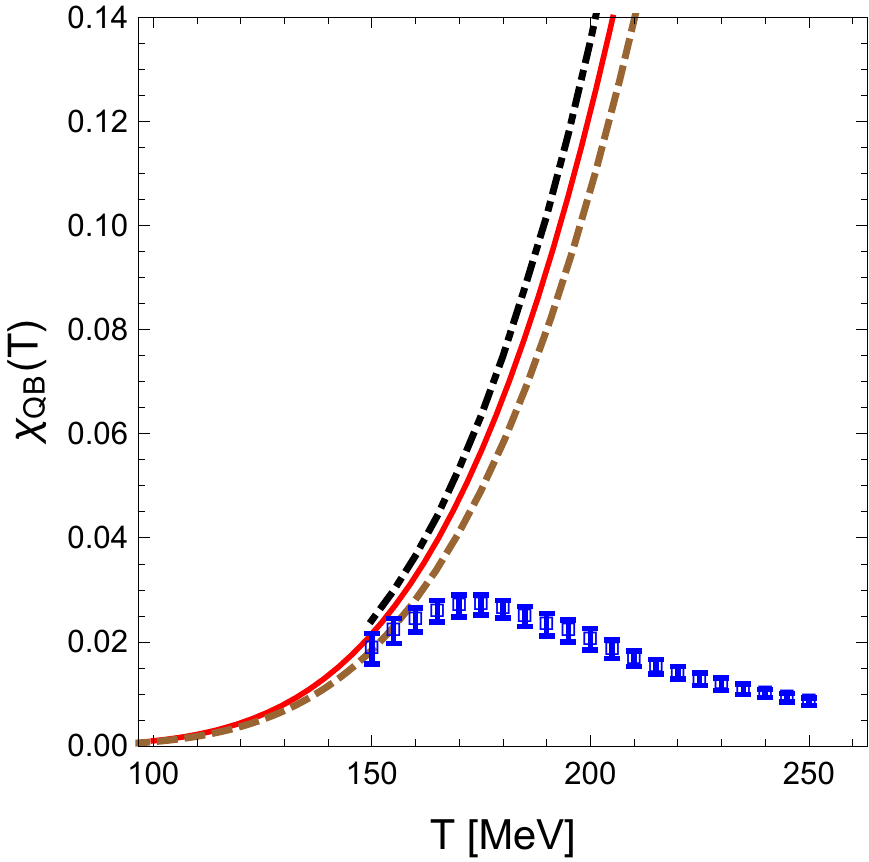} &
 \includegraphics[width=0.26\textwidth]{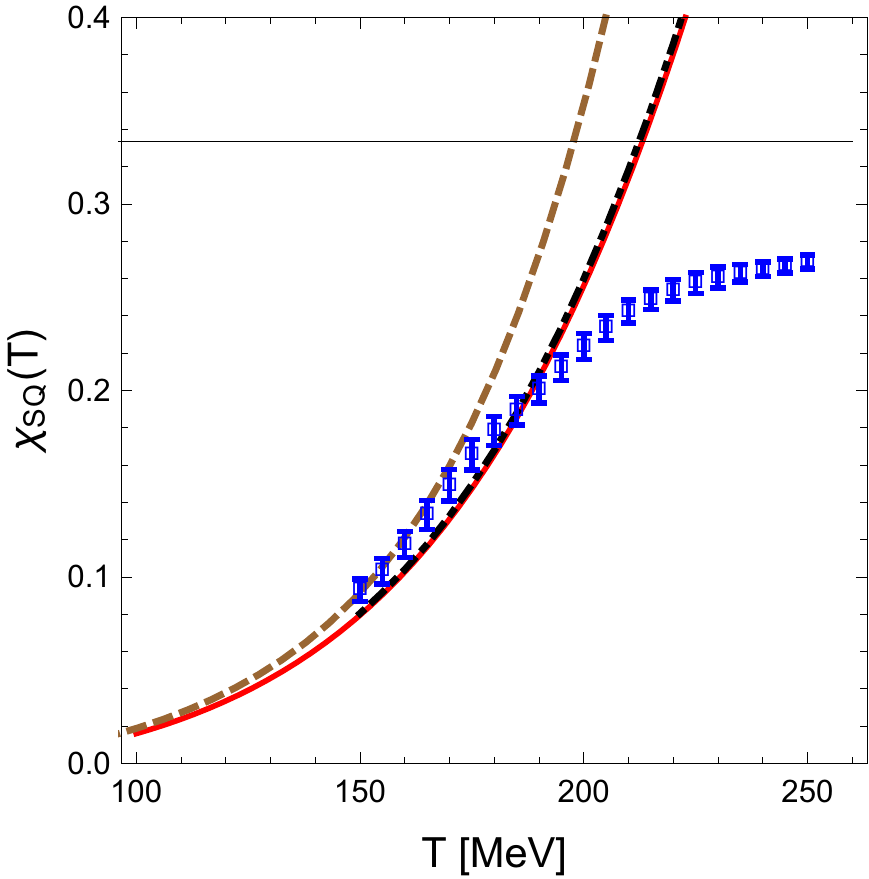}
\end{tabular}
\vspace{-0.4cm}
 \caption{\it Baryon, charge and strangeness susceptibilities from HRG
   with the spectrum of: i) PDG, ii) PDG including a Gaussian width profile in the resonances (PDG($\Gamma$)), and iii) RQM. We display as dots the HotQCD~\cite{Bazavov:2012jq} lattice data.}
\label{fig:Chi2}
\end{figure*}

Within the HRG approach, the charges are carried by various species of hadrons, $Q_a = \sum_i q^i_a N_i$, where $q_a^i \in \{ Q_i , B_i , S_i \}$., and $N_i$ is the number of hadrons of type~$i$. By using in Eq.~(\ref{eq:susc}) the thermodynamic potential of this model, cf. Eq.~(\ref{eq:Zhrgm}), one gets
\begin{eqnarray}
\hspace{-1cm} &&\chi_{ab}(T) =   \label{eq:chi_HRGM} \\
\hspace{-1cm} &&\quad = \frac{1}{2\pi^2}  \hspace{-0.2cm} \sum_{i \in {\rm Hadrons}}  \sum_{k=1}^\infty  \zeta_i^{1+k} g_i q_i^a q_i^b \left( \frac{M_i}{T} \right)^2 K_2\left( k \frac{M_i}{T} \right) \,,   \nonumber
\end{eqnarray}
where $K_2(z)$ refers to the Bessel function of the second kind. Using this formula, we have computed the susceptibilities in the three different sectors,~$Q_a \in \{ Q, B, S\}$. The results, compared to the lattice data of~\cite{Bazavov:2012jq}, are plotted in Fig.~\ref{fig:Chi2}. We conclude that, in general, there is a good description of lattice data for $T \lesssim 150$ MeV. However, the agreement for some of the susceptibilities is not as good as for the EoS, leading to the conclusion that fluctuations may serve as a diagnostic tool to study missing states in the spectrum in different flavor sectors~\cite{RuizArriola:2016qpb}.

\section{Correlations in a thermal medium}
\label{sec:correlations}

While the fluctuations are global quantities, in this section we will extend the above results to local correlators of conserved quantities. These depend on the spin of the particles. We will study in details the case of spin~$1/2$, and present results that include the contribution of particles up to much higher spin.

\subsection{Free particles of spin $1/2$}
\label{subsec:spin12}

The Lagrangian density for Dirac fermions in Euclidean space is
\begin{equation}
{\cal L}(x) = \sum_{a=1}^{N_f} \overline{\Psi}_a(x) (\slash \!\!\!\partial + m_a) \Psi_a(x) \,,
\end{equation}
where $m$ is the mass and $\slash \!\!\!\partial = \gamma^\mu \partial_\mu$. The partition function of the system is
\begin{equation}
Z = \int {\cal D}\Psi(x) {\cal D}\overline{\Psi}(x) e^{-\int_0^\beta dx_0 \int d^3x {\cal L}(x) }  \,,
\end{equation}
where it is considered that fermions are antiperiodic in Euclidean time, $\Psi(x_0+\beta,\vec{x}) = -\Psi(x_0,\vec{x})$ with $\beta=1/T$ the inverse of temperature. The vector currents are defined as
\begin{equation}
j^{\mu}_a(x) = \overline{\Psi}(x) \gamma^\mu \hat{Q}_a \Psi(x) \,, \qquad a = 1, \cdots , N_f \,,
\end{equation}
where~$\hat{Q}_a$ is a matrix that specifies the charge. The temporal components of the currents are the conserved charges, i.e.~$\rho_a(x) \equiv j_a^0(x)$. In the following we will be interested in the retarded correlators~\footnote{In absence of anomalies, the correlators fulfill the conservation equation~$\partial_\mu \langle j^\mu_a(x) j^\nu_b(0)\rangle = 0$.}
\begin{equation}
C_{ab}^{\mu\nu}(x) = \langle j^\mu_a(x) j^\nu_b(0)\rangle  \,.
\end{equation}
Given the propagator of fermions of spin $1/2$ in position space
\begin{equation}
 S_{1/2}(x) = - \int \frac{d^4k}{(2\pi)^4} \frac{i {\slash \!\!\!k} + m}{k^2+m^2} e^{-i k x} \,, \label{eq:S12}
\end{equation}
the correlator writes
\begin{equation}
\langle j_a^\mu(x) j_b^\nu(0)\rangle = \delta_{ab} \langle S_{1/2}(x) \gamma^\mu S_{1/2}(-x) \gamma^\nu \rangle  \,.
\end{equation}
After a straightforward computation, the result for the correlator at zero temperature is
\begin{eqnarray}
&&\hspace{-1cm}\langle j^\mu(x) j^\nu(0)\rangle = 4 \Big[ 2 (\partial^\mu \Delta(x))  (\partial^\nu \Delta(x)) \nonumber \\
&&\qquad - ((\partial_\alpha \Delta(x))^2 + m^2 \Delta(x)^2) \eta^{\mu\nu} \Big] \,, \label{eq:corrT0}
\end{eqnarray}
where
\begin{equation}
\Delta(x) = \int \frac{d^4k}{(2\pi)^4} \frac{e^{-i k_\mu x^\mu}}{k^2 + m^2}  = \frac{m}{4\pi^2} \frac{K_1(m|x|)}{|x|} \,,  \label{eq:DeltaT0}
\end{equation}
with $|x| = \sqrt{x_0^2 + \vec{x}^2}$. The explicit result for the correlator at zero temperature is then
\begin{eqnarray}
&&\hspace{-1.5cm}\langle j^\mu(x) j^\nu(0)\rangle = \frac{4m^4}{(4\pi^2)^2} \Bigg[ \left( \frac{K_2(m|x|)}{|x|^2} \right)^2 \!\!\left[ 2 x^\mu x^\nu - \eta^{\mu\nu} x^2 \right] \nonumber \\
&&\hspace{0.5cm} -\left( \frac{K_1(m|x|)}{|x|} \right)^2 \eta^{\mu\nu} \Bigg] \,.
\end{eqnarray}
The static part of this correlator has the following behavior at small distances
\begin{equation}
\langle j_a^0(\vec{x}) j_b^0(0)\rangle \simeq \delta_{ab} \left( -\frac{1}{\pi^4 r^6} + \frac{m^2}{4\pi^4 r^4} + {\cal O}(r^{-2}) \right) \,, 
\end{equation}
with~$r = |\vec{x}|$.

We can extend straightforwardly this result to finite temperature. By using the Poisson's summation formula, one gets the following rule to transform zero to finite temperature expressions~\cite{Megias:2004hj}
\begin{equation}
\hspace{-0.6cm}  \int \frac{dk_0}{2\pi} F(k_0,\vec{k}) \to i \sum_{n=-\infty}^{\infty}\zeta^n \int \frac{dk_4}{2\pi} F(ik_4,\vec{k}) e^{ink_4/T} \,.
\end{equation}
This formula, applied to Eq.~(\ref{eq:DeltaT0}), leads to a summation over the number of thermal loops: $n=0$ corresponds to the zero temperature contribution, while $n\ne 0$ are finite temperature corrections. Then, the correlator at finite temperature writes as in Eq.~(\ref{eq:corrT0}) with
\begin{equation}
\Delta(x) \to \Delta_T(x) = \frac{m}{4\pi^2} \sum_{n=-\infty}^{+\infty} \zeta^n \frac{K_1(m|x|)}{|x|} \,,
\end{equation}
where $|x| = \sqrt{\left(x_0 - n/T \right)^2 + \vec{x}^2}$. Finally, from a comparison of the zero and finite temperature correlators, one finds that the thermal corrections in the static correlator at small distances read
\begin{eqnarray}
&&\hspace{-1.4cm} \langle j^0_a(\vec{x}) j^0_b(0)\rangle_T - \langle j^0_a(\vec{x}) j^0_b(0)\rangle_{T=0} = \nonumber \\
&&\hspace{-0.3cm} = \frac{\delta_{ab}}{r^2} \frac{m^2 T}{\pi^2} \left( m K_{1}\left(\frac{m}{T}\right) + 2T  K_{2}\left(\frac{m}{T}\right) \right) + \cdots \,,
\end{eqnarray}
so that in the case of spin $1/2$ particles the corrections start at ${\cal O}(r^{-2})$. 

\begin{figure*}[htb]
\centering
 \begin{tabular}{c@{\hspace{4.5em}}c@{\hspace{4.5em}}c}
 \includegraphics[width=0.26\textwidth]{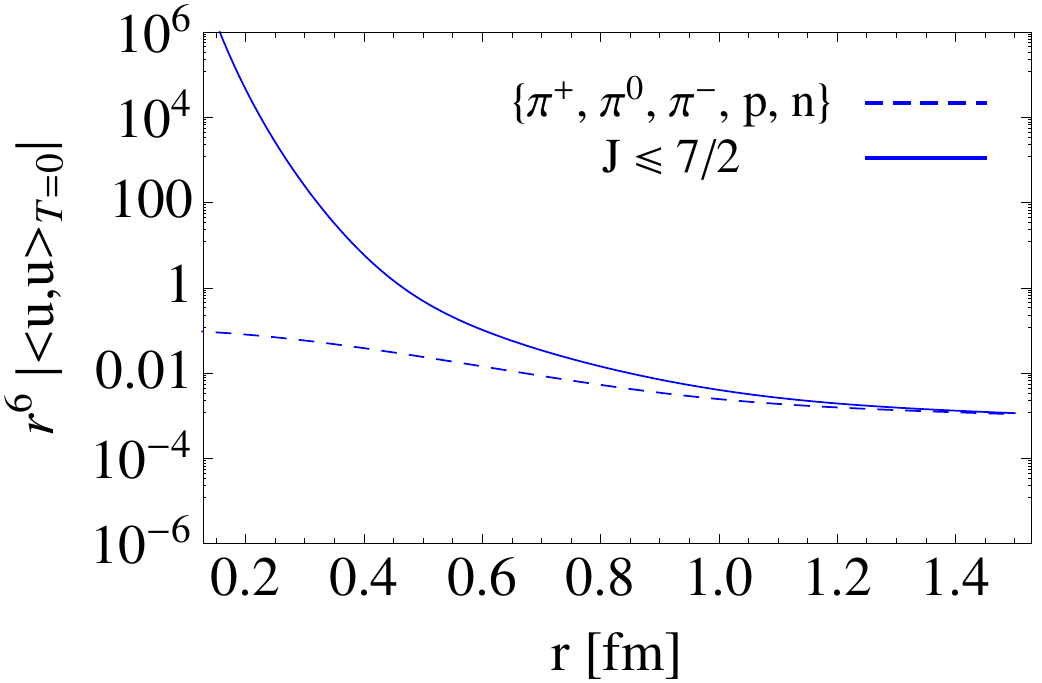} &
 \includegraphics[width=0.26\textwidth]{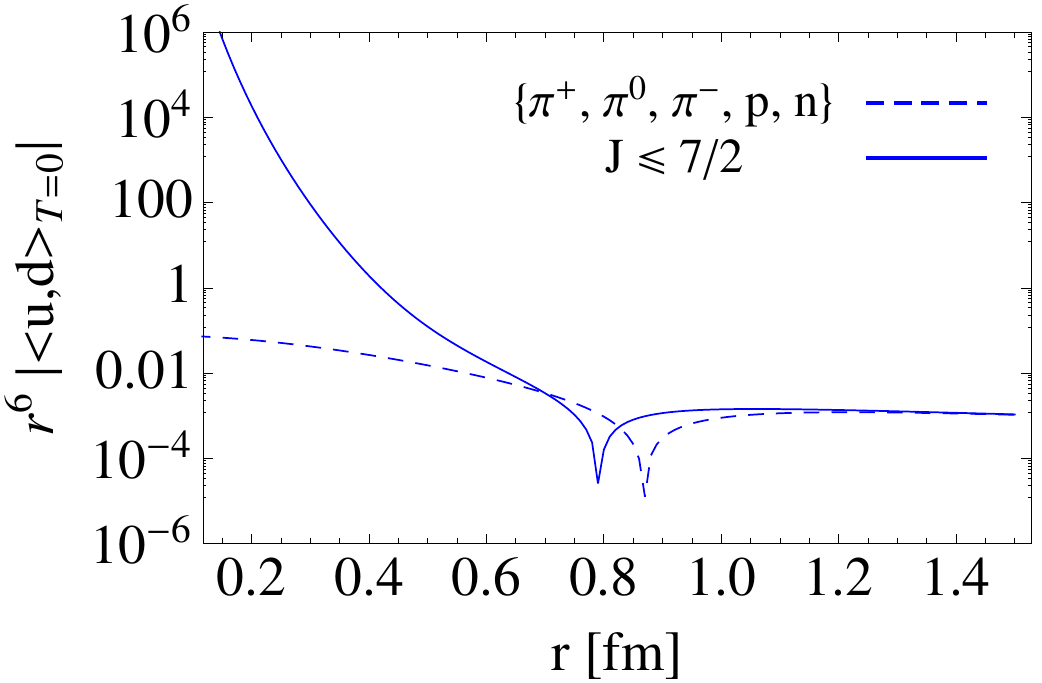} &
 \includegraphics[width=0.26\textwidth]{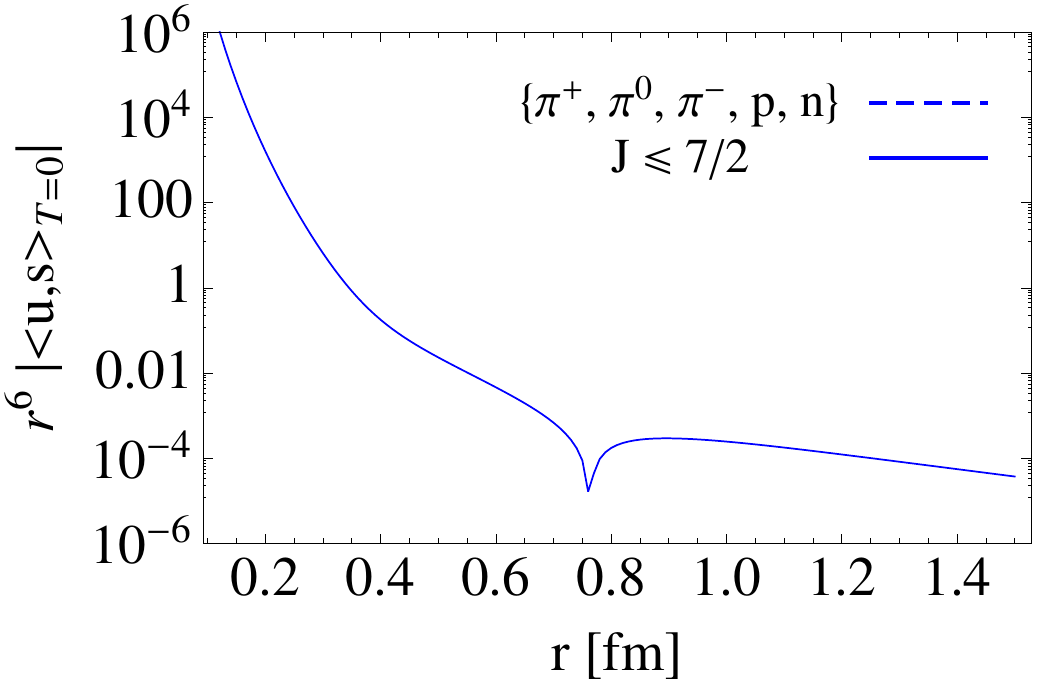} \\
 \includegraphics[width=0.26\textwidth]{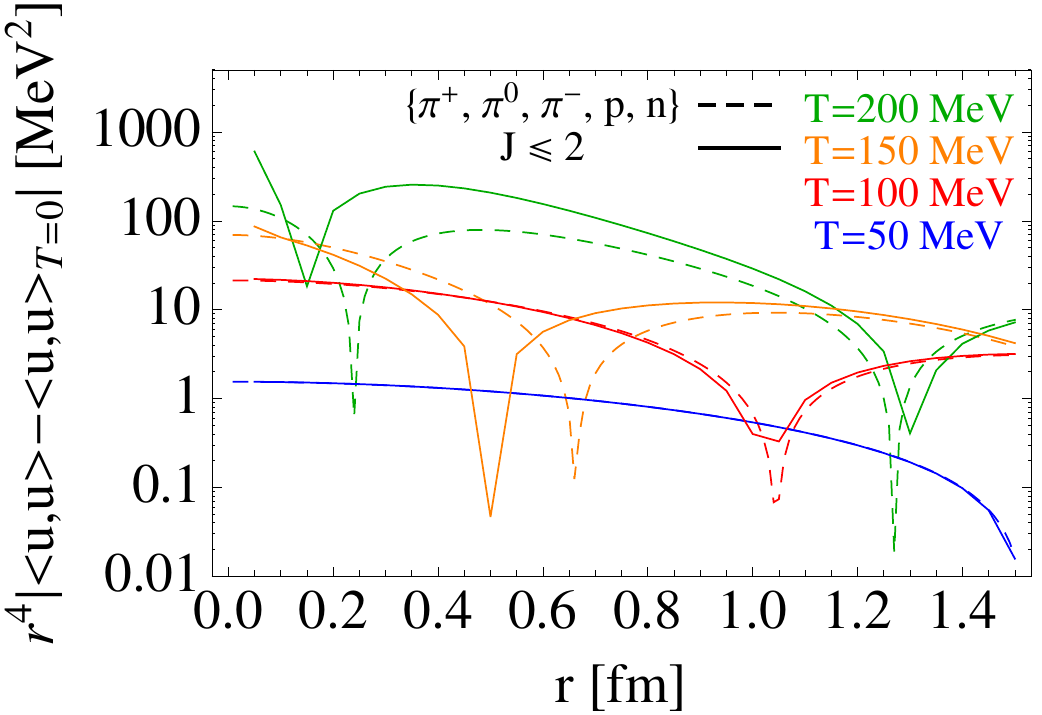} &
 \includegraphics[width=0.26\textwidth]{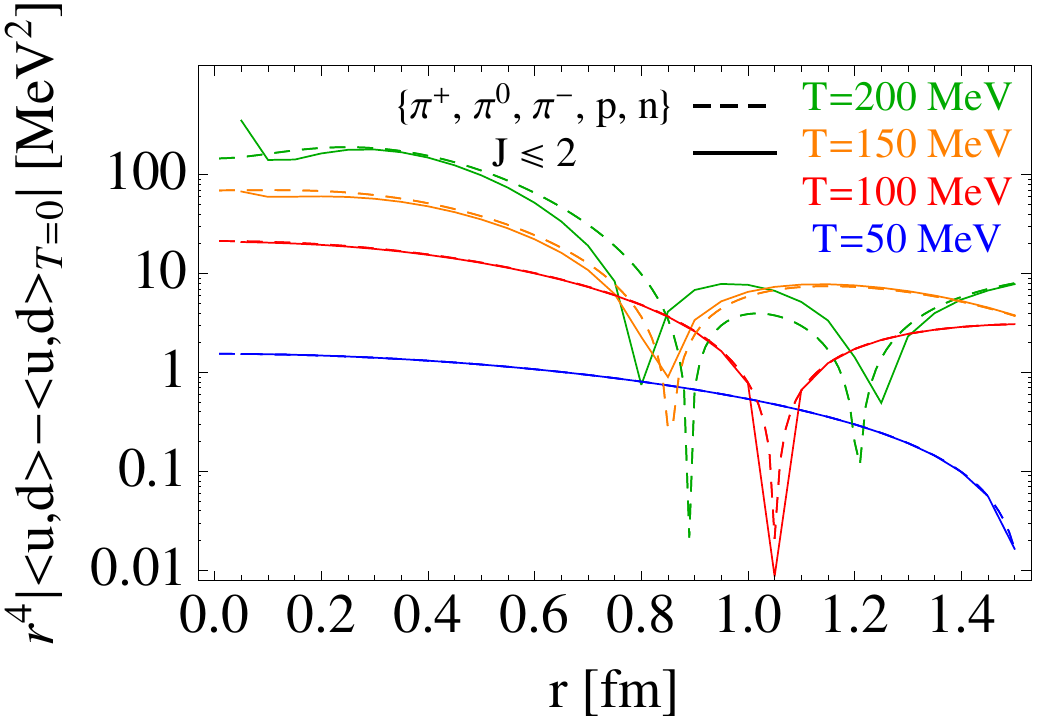} &
 \includegraphics[width=0.26\textwidth]{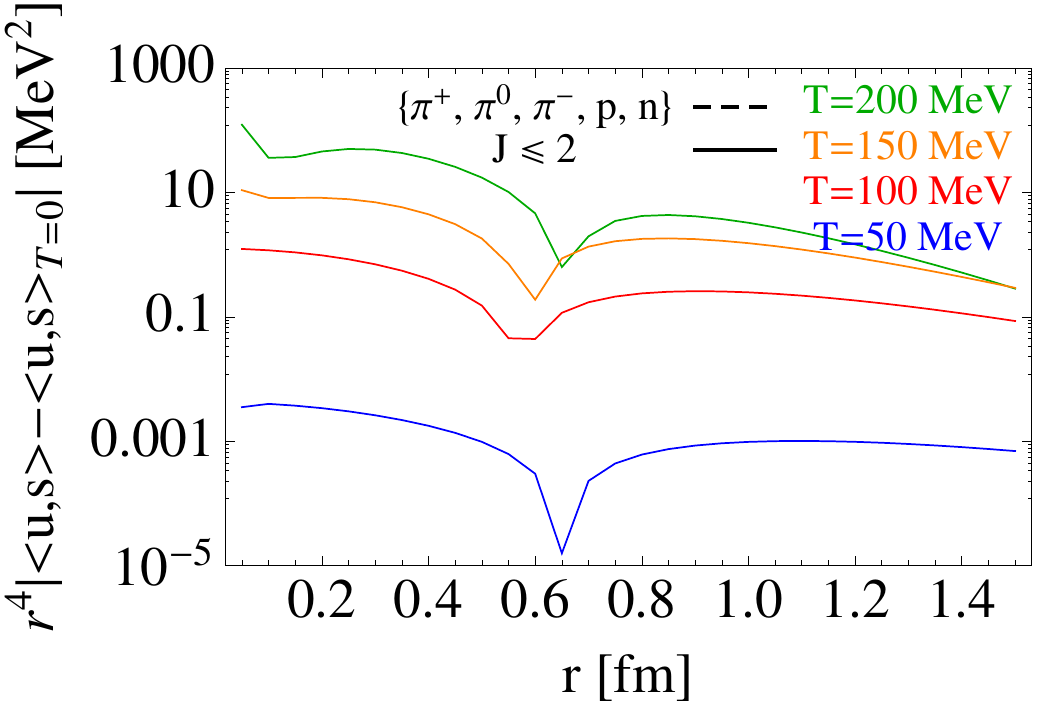}
\end{tabular}
\vspace{-0.4cm}
 \caption{\it Static correlators $C_{00}(r)$ at zero temperature (upper panels) and finite temperature (lower panels), including pions and nucleons (dashed lines) and hadrons with $J \le 7/2$ and $J \le 2$ from the RQM spectrum~\cite{Godfrey:1985xj,Capstick:1986bm} (solid lines).
  }
\label{fig:correlators}
\end{figure*}

\subsection{Free particles of any spin}

It is possible to extend the above analysis to particles of any spin by using the Bargmann-Wigner formalism~\cite{Bargmann:1948ck} (details will be provided elsewhere~\cite{Megias:2018inprogress}). Let us mention that the small distance behavior of the static correlators, either at zero or finite temperature, reads (for $J>0$ and up to a factor)
\begin{equation}
C^{ab\, J}_{00}(r) \stackrel[r \to 0]{\sim}{} \delta_{ab}
\frac{m^2}{r^4} \frac{1}{(mr)^{4J}}
\,, \label{eq:correJ_smalldistance}
\end{equation}
while for $J=0$ it is~$C^{ab\, J=0}_{00}(r) \stackrel[r \to 0]{\sim}{} \delta_{ab}
\frac{1}{r^6}$.

\subsection{Correlations in the HRG model}

Within the HRG model the correlator writes
\begin{equation}
\hspace{-0.3cm} C^{ab}_{\mu\nu}(x) \equiv \langle j^a_\mu(x) j^b_\nu(0) \rangle = \sum_{i \in {\rm Hadrons}} \frac{1}{2} q^a_i q^b_i C^{ab\, J_i}_{\mu\nu}(x) \,, \label{eq:Cab_HRGM1}
\end{equation}
where~$ q^a_i \in \{ Q_i, B_i , S_i\}$, and~$C^{ab\, J}_{\mu\nu}(x)$ are the correlators of free particles of spin~$J$. The index $i$ in this formula stands for any hadron, distinguishing between spin $J_i$, isospin and particle-antiparticle. As an example, the lowest lying states in the meson and hadron spectrum corresponding to pions and protons/neutrons are
\begin{equation}
\hspace{-0.6cm} i \in \{ \pi^+ , \pi^0 , \pi^- , p\uparrow, p\downarrow, \bar{p}\uparrow, \bar{p}\downarrow, n\uparrow, n\downarrow, \bar{n}\uparrow, \bar{n}\downarrow  \}.
\end{equation}
An expression equivalent to Eq.~(\ref{eq:Cab_HRGM1}) is
\begin{eqnarray}
 C^{ab}_{\mu\nu}(x) &=&  \sum_{M \in {\rm Mesons}} \!\!\! \frac{1}{2}(2J_M+1) q^a_M q^b_M C_{\mu\nu}^{J_M}(x) \nonumber \\
&+& \sum_{B \in {\rm Baryons} > 0} \!\!\!\!\!\! (2J_B+1) q^a_B q^b_B C_{\mu\nu}^{J_B}(x) \,, \label{eq:Cab_HRGM2}
\end{eqnarray}
where~$M$ and $B$ run now over the spin multiplets of mesons and baryons, each of them with degeneracy $(2J_M+1)$ and $(2J_B+1)$. Baryons and antibaryons contribute to the correlators in the same amount, so that we have considered in $\sum_B$ a summation over baryons only, and multiplied it by a factor~$2$. The lowest lying states contributing to Eq.~(\ref{eq:Cab_HRGM2}) are $M \in \{ \pi^+ , \pi^- , \pi^0\}$ and $B \in \{ p , n\}$. We display in Fig.~\ref{fig:correlators} the result of the static correlators at zero and finite temperature within the HRG approach.

The static version of these correlators are related to the susceptibilities~$\chi_{ab}(T)$ through
\begin{equation}
\chi_{ab}(T) = \frac{1}{T^3} \int d^3x \, C_{00}^{ab}(0,\vec{x}) \,.
\end{equation}
This formula applies for particles of any spin~$J$, and so it is also valid for the susceptibilities and correlators within the HRG approach, cf. Eqs.~(\ref{eq:chi_HRGM}) and (\ref{eq:Cab_HRGM2}). From the asymptotic behaviors of these two quantities
\begin{equation}
\chi_{ab}(T) \stackbin[T \to 0]{}{\sim} e^{-m/T} \quad \textrm{and} \quad C_{00}^{ab}(0,\vec{x}) \stackbin[r \to \infty]{}{\sim} e^{-2 m r} \,,
\end{equation}
where~$m$ is the mass of the lowest-lying state, we may conjecture the
existence of a formal analogy between the static correlators at zero
temperature and the finite temperature susceptibilities, after
considering the replacement $r \leftrightarrow 1/T$. A natural
consequence of this analogy is that the existence of a limiting
temperature for the validity of the hadronic representation of
$\chi_{ab}(T)$, i.e. $T < T_H$, cf. Eq.~(\ref{eq:ZHRG}), might have
its counterpart in the existence of a limiting distance in the
hadronic representation of the correlators $(r >
r_H)$~\cite{Megias:2018pwh}. In fact, given the behavior of the static
correlators for particles of spin $J$,
cf. Eq.~(\ref{eq:correJ_smalldistance}), we find that after summation
over hadrons of higher and higher spin, the static correlators within
the HRG model present a divergence at some finite value of the
distance
\begin{equation}
C_{00}^{\HRG}(r) = \sum_J C_{00}^J(r)  \stackrel[r \to r_H^+]{\longrightarrow}{} \frac{K}{r_H - r} \,,
\end{equation}
when considering a degeneracy of states of the form
\begin{equation}
g_J = e^{c J}  \,, \quad \textrm{with} \qquad c = 4 \log(m r_H) \,.
\end{equation}
Here~$r_H$ can be interpreted as a Hagedorn distance, in analogy to
the Hagedorn temperature $T_H$ at finite $T$. When using $T_H \approx
150 \MeV$ one gets $r_H \simeq 1/(2\pi T_H) \approx 0.21 \fm$, where
the factor $1/(2\pi)$ is standard in finite $T$ computations. Of course, this is just a consequence of stretching the domain of validity of the HRG; at short distances quark degrees of freedom set in.

\section{Conclusions}
\label{sec:conclusions}

In the present work we have studied the EoS, fluctuations and correlations of conserved charges in a thermal medium by using the HRG approach. Being the predictions of this model sensitive to the spectrum of QCD, a comparison with lattice data serves as a diagnostic tool to study the possible existence of missing states in the hadron spectrum. This analysis is also related to the question of the completeness of the QCD spectrum. From this comparison, we conclude that the HRG model works at $T \lesssim 0.8 T_c$, and in this regime of temperatures hadrons can be considered as a complete basis of states. However, close to $T_c$ many hadrons are needed to saturate the sum rules, leading to the question of what states are needed when approaching $T_c$ from below. While the EoS is sensitive to the spectrum of QCD as a whole, fluctuations and correlations of conserved charges in the confined phase of QCD allow to study missing states in three different sectors: i) electric charge, ii) baryon number, and iii) strangeness.  Our results for the fluctuations fairly agree with the lattice results for temperatures~$T \lesssim 150 \,\MeV$. However, to our knowledge there are no lattice data for the correlations, and we expect that a comparison of our predictions with future lattice data will help in the study of missing states. Finally, let us mention that similar studies for other thermal observables show indications of the existence of exotic states. In particular, the Polyakov loop and Entropy shift due to a heavy quark in the medium suggest that there are in the QCD spectrum: i) conventional missing states~($[Q \bar q]$ and $[Qqq]$), and ii) hybrid states ($[Q\bar q g]$ and $[Qqqg]$)~\cite{Megias:2012kb,Megias:2016onb,RuizArriola:2016qpb}. All these studies constitute new tools to shed some light on the present unsolved problems about the spectroscopy of the QCD, including the existence of exotic states.

\section*{Acknowledgements} 
This work is supported by the Spanish MINECO and European FEDER funds (grants
FIS2014-59386-P, FIS2017-85053-C2-1-P and FPA2015-64041-C2-1-P), Junta
de Andaluc\'{\i}a (grant FQM-225) and Basque Government (grant
IT979-16). The research of E.M. is also supported by the Ram\'on y
Cajal Program of the Spanish MINECO, and by the Universidad del
Pa\'{\i}s Vasco UPV/EHU, Bilbao, Spain, as a Visiting Professor.


\end{document}